\pdfoutput=1
\documentclass[10pt,conference]{IEEEtran}
\IEEEoverridecommandlockouts
\usepackage{cite}
\usepackage{amsmath,amssymb,amsfonts}
\usepackage{graphicx}
\usepackage{textcomp}
\usepackage{xcolor}
\def\BibTeX{{\rm B\kern-.05em{\sc i\kern-.025em b}\kern-.08em
    T\kern-.1667em\lower.7ex\hbox{E}\kern-.125emX}}
\usepackage{framed}
\usepackage{listings}
\usepackage{xcolor}
\usepackage{color}
\usepackage{multirow}
\usepackage{algorithm}

\usepackage{amsmath}
\usepackage{amssymb}
\usepackage{threeparttable}
\usepackage{url}
\usepackage[most]{tcolorbox}
\usepackage{hyperref}
\usepackage[hyphenbreaks]{breakurl}
\usepackage{algpseudocode}
\usepackage{colortbl}
\usepackage{booktabs}
\usepackage{balance}
\usepackage{subcaption}
\hypersetup{
    colorlinks = true,
    linkcolor=blue,
    filecolor=blue,      
    urlcolor=blue,
    citecolor=cyan,
}
\usepackage{xspace}
\usepackage{graphicx}
\usepackage{fvextra}

\usepackage[shortlabels]{enumitem}
\setlist[enumerate]{nosep}

\definecolor{codegreen}{rgb}{0,0.6,0}
\definecolor{codegray}{rgb}{0.5,0.5,0.5}
\definecolor{codepurple}{rgb}{0.58,0,0.82}
\definecolor{backcolour}{rgb}{0.95,0.95,0.92}
\definecolor{lightgreen}{rgb}{0,0.4,0}
\definecolor{lightred}{rgb}{0.4,0,0}

\definecolor{mygray}{gray}{.9}

\lstdefinestyle{mystyle}{
    commentstyle=\color{brown},
    keywordstyle=\color{magenta},
    numberstyle=\tiny\color{codegray},
    stringstyle=\color{codepurple},
    basicstyle=\ttfamily\footnotesize,
    breakatwhitespace=false,         
    breaklines=true,                 
    captionpos=b,                    
    keepspaces=true,                 
    numbers=none,                    
    numbersep=5pt,                  
    showspaces=false,                
    showstringspaces=false,
    showtabs=false,                  
    tabsize=2,
    escapeinside={<@}{@>}
}

\newcommand{\tool}{\textsc{TypeGen}\xspace}
\newcommand{\typepy}{Type4Py\xspace}
\newcommand{\typebert}{TypeBERT\xspace}
\newcommand{\hityper}{HiTyper\xspace}
\newcommand{\codet}{CodeT5\xspace}
\newcommand{\gptneo}{GPT-Neo\xspace}

\newcommand{\gpt}{GPT-3.5\xspace}
\newcommand{\gptj}{GPT-J\xspace}
\newcommand{\typewriter}{TypeWriter\xspace}
\newcommand{\incoder}{InCoder\xspace}
\newcommand{\unixcoder}{UniXcoder\xspace}
\newcommand{\chatgpt}{ChatGPT\xspace}
\newcommand{\codegen}{CodeGen\xspace}

\newcommand\etal{{\it{et al.\ }}}

\newcommand{\tabincell}[2]{\begin{tabular}{@{}#1@{}}#2\end{tabular}}
\newcommand{\eat}[1]{\if 0 #1 \fi}

\newcommand{\g}{\cellcolor{mygray}}

\newfont{\mycrnotice}{ptmr8t at 7pt}
\newfont{\myconfname}{ptmri8t at 7pt}

\newcommand{\answer}[2]{
\begin{tcolorbox}[breakable,width=\linewidth,boxrule=0pt,top=1pt, bottom=1pt, left=1pt,right=1pt, colback=gray!20,colframe=gray!20]
\textbf{Answer to RQ#1:} #2
\end{tcolorbox}
}
\begin{document}

\title{Generative Type Inference for Python}

\author{
    \IEEEauthorblockN{Yun Peng\IEEEauthorrefmark{2}, Chaozheng Wang\IEEEauthorrefmark{2}, Wenxuan Wang\IEEEauthorrefmark{2}, Cuiyun Gao\IEEEauthorrefmark{3}\thanks{* Cuiyun Gao is the corresponding author.}, Michael R. Lyu\IEEEauthorrefmark{2}}
    \IEEEauthorblockA{\IEEEauthorrefmark{2} Computer Science and Engineering Department, The Chinese University of Hong Kong, Hong Kong, China}
    \IEEEauthorblockA{\IEEEauthorrefmark{3} School of Computer Science and Technology, Harbin Institute of Technology, Shenzhen, China}
    \IEEEauthorblockA{\{ypeng, wxwang, lyu\}@cse.cuhk.edu.hk, adf111178@gmail.com, gaocuiyun@hit.edu.cn}
}

\maketitle

\begin{abstract}
Python is a popular dynamic programming language, evidenced by its ranking as the second most commonly used language on GitHub. However, its dynamic type system can lead to potential type errors, leading researchers to explore automatic type inference approaches for Python programs. Existing type inference approaches can be generally grouped into three categories, i.e., rule-based, supervised, and cloze-style approaches. The rule-based type inference approaches can ensure the accuracy of predicted variable types, but they suffer from low coverage problems caused by dynamic features and external calls. Supervised type inference approaches, while feature-agnostic and able to mitigate the low coverage problem, require large, high-quality annotated datasets and are limited to pre-defined types. As zero-shot approaches, the cloze-style approaches reformulate the type inference problem into a fill-in-the-blank problem by leveraging the general knowledge in powerful pre-trained code models. However, their performance is limited since they ignore the domain knowledge from static typing rules which reflect the inference logic. What is more, their predictions are not interpretable, hindering developers' understanding and verification of the results.

This paper introduces \tool, a few-shot generative type inference approach that incorporates static domain knowledge from static analysis. \tool creates chain-of-thought (COT) prompts by translating the type inference steps of static analysis into prompts based on the type dependency graphs (TDGs), enabling language models to learn from how static analysis infers types. By combining COT prompts with code slices and type hints, \tool constructs example prompts from human annotations. \tool only requires very few annotated examples to teach language models to generate similar COT prompts via in-context learning. Moreover, \tool enhances the interpretability of results through the use of the \textit{input-explanation-output} strategy, which generates both explanations and type predictions in COT prompts. Experiments show that \tool outperforms the best baseline \typepy by 10.0\% for argument type prediction and 22.5\% in return value type prediction in terms of top-1 Exact Match by using only five examples.  Furthermore, \tool achieves substantial improvements of 27\% to 84\% compared to the zero-shot performance of large language models with parameter sizes ranging from 1.3B to 175B in terms of top-1 Exact Match.
\end{abstract}

\begin{IEEEkeywords}
type inference, chain-of-thought, generative model
\end{IEEEkeywords}

\section{Introduction}\label{sec:intro}

With the boom of artificial intelligence and data science, Python is becoming increasingly popular in recent years. As a dynamically typed programming language, Python is famous for its convenience and usability. The dynamic type system makes it possible to reuse the same code snippets for different functionalities, which significantly improves development efficiency.  However, this convenience comes with a cost. The dynamic type system poses a threat to the reliability of Python software by introducing more type errors. Oh \etal~\cite{pyter} find that 30\% of questions raised by developers at GitHub and Stack Overflow are related to type errors. To reduce potential type errors, Python Software Foundation introduces type annotations in a series of Python Enhancement Proposals (PEPs)~\cite{pep484,pep544,pep585,pep589}. Manually annotating types for each variable in Python programs is overwhelming, so many automatic type inference approaches~\cite{typilus, pradel20typewriter, type4py,peng22hityper} are proposed to infer types statically to release the burden of developers. Automatic type inference approaches work along with static type checkers~\cite{pytype,pyre,pyright,mypy} to detect potential type errors for Python programs~\cite{pyter,peng2023domain}.

The earliest proposed automatic type inference approaches are rule-based, as described in previous work~\cite{javainfer05,jsinfer09,rubyinfer09,emmi16abstracttypeinfer,Zvonimir21dataflowtypeinfer,sheng16principaltypeinfer}. These approaches rely on pre-defined typing rules and static analysis to accurately infer types. However, they are limited by the low coverage problem since the types of many variables in Python programs cannot be resolved statically.

Inspired by the remarkable achievements of deep learning in the natural language processing (NLP) field, supervised type inference approaches~\cite{typilus,pradel20typewriter,type4py,wei2020lambdanet,jesse21typebert} take the code context of the target variable as input and leverage deep learning models to classify the context into one type. They can naturally avoid the low coverage problem of rule-based approaches, as deep learning models are feature-agnostic and make predictions based on probability rather than rules. Taking advantage of identifiers in code, supervised type inference approaches are quite effective after training on a large dataset of annotated code. Despite the effectiveness, supervised type inference approaches based on classification methods classify inputs into pre-defined type categories and perform badly on rare types. Peng~\etal~\cite{hityper} propose a hybrid type inference approach \hityper to mitigate these problems by using deep learning models to recommend type predictions for static analysis. However, deep learning models in supervised approaches require large high-quality datasets of type annotations, which needs substantial human efforts.

Cloze-style type inference approaches~\cite{feng2020codebert, guo21graphcodebert, wang21codet5, guo22unixcoder, daniel22incoder} transform the type inference problem into a fill-in-the-blank problem by adding masks on the locations of type annotations in code. These approaches are well-aligned with the pre-training objectives of pre-trained code models and do not require large datasets, making them suitable for zero-shot settings.  However, they face the following challenges:

\begin{figure*}[t]
    \centering
    \includegraphics[width = 1.0\textwidth]{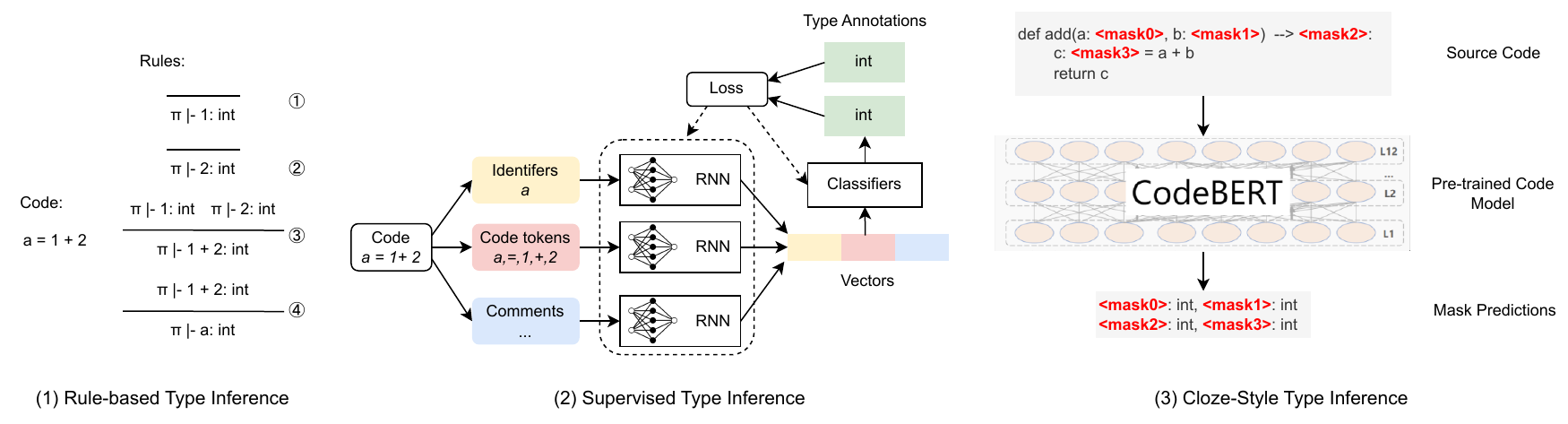}
    \caption{Three kinds of type inference approaches.}
    \label{fig:background}
\end{figure*}

\textit{1) Lack of static domain knowledge.} Cloze-style approaches are characterized by the insertion of masks in source code, allowing pre-trained code models to predict the missing type information. While they have the advantage of not requiring large datasets, unlike supervised approaches, they rely solely on the general knowledge that pre-trained code models acquire during the pre-training phase. Consequently, their performance may be suboptimal, as they lack an understanding of how types are constructed based on typing rules.

\textit{2) Lack of interpretability.} Current learning-based type inference approaches, including supervised and cloze-style approaches, adopt the \textit{input-output} methodology, taking code as input and outputting single types. However, they provide no idea about how deep learning models reach the \textit{output} types from the \textit{input} code. This lack of transparency makes it challenging for developers to comprehend and validate the predicted types, particularly when there are insufficient static constraints.

\textbf{Our work.} We propose \tool, the first few-shot generative type inference approach for Python programs. \tool has four phases, including code slicing, type hint collection, chain-of-thought (COT) prompt construction, and type generation. In the code slicing phase, \tool generates type dependency graphs (TDGs) and builds code slices based on TDG as the contexts of target variables, i.e., the variables whose types need to be inferred. In the type hints collection phase, \tool collects all available user-defined types and third-party types as type hints via import analysis to provide additional knowledge that does not exist in code slices. In the COT prompt construction phase, \tool translates the inference steps of static analysis for target variables into a COT prompts~\cite{cot}. The code slices, type hints and COT prompts generated in the first three phases are combined as the \textit{example prompts}, which provides rich static domain knowledge. In the last type generation phase, \tool adopts the \textit{in-context learning} (ICL) methodology and constructs the input prompt by concatenating several example prompts and the \textit{target variable prompt}, which includes code slice as well as type hints of the target variable.
A language model is then invoked to complete the input prompt with the COT prompt of the target variable. With both explanations and predicted types in the generated COT prompts, \tool can improve the interpretability of results.

We evaluate \tool on the widely-used ManyTypes4Py dataset~\cite{type4pydataset}. Our experiment results show that \tool outperforms the most advanced baseline \typepy~\cite{type4py} by 10.0\% for argument types and 22.5\% for return value types in terms of top-1 Exact Match. Furthermore, we observe that \tool can achieve improvements of 27\% $\sim$ 84\% over the zero-shot performance of language models with parameter sizes ranging from 1.3B to 175B in terms of top-1 Exact Match, which are $2\times \sim 3\times$ of the improvements achieved by the standard ICL method without static domain knowledge.

\textbf{Contributions.} We summarize our contributions as follows.
\begin{itemize}
    \item To the best of our knowledge, we propose the first few-shot generative type inference approach named \tool for Python.
    \item We propose a novel prompt design to incorporate different static domain knowledge into language models, which includes code slices, type hints, and COT prompts.
    \item Extensive experiments demonstrate the effectiveness of \tool compared with supervised and cloze-style type inference approaches, as well as the capability of \tool on language models with different parameter sizes.
\end{itemize}

\section{Background and Related Work}\label{sec:background}

We classify existing type inference approaches into three categories: rule-based, supervised and cloze-style approaches, and present an overview of three kinds of type inference approaches in Fig.~\ref{fig:background}.

\subsection{Rule-based Type Inference}

Rule-based approaches for type inference rely on predefined rules to determine the types of variables. Fig.~\ref{fig:background}(1) shows an example where four rules are associated with the type inference of variable \textit{a}. Each rule has premises (above the line) and conclusions (below the line). A rule can be triggered only if all premises are known, and then the result type is given based on the conclusion.

To address the need for static type hints in dynamically typed programming languages, various approaches have been proposed for type inference and checking, such as Pyright~\cite{pyright} and Pylance from Microsoft, Pyre from Meta~\cite{pyre}, Pytype from Google~\cite{pytype}, and Python's official type checker mypy~\cite{mypy}. In addition to industry tools, some academic approaches have been proposed for type inference in different programming languages, such as Python and JavaScript~\cite{javainfer05,jsinfer09,rubyinfer09,emmi16abstracttypeinfer,Zvonimir21dataflowtypeinfer,sheng16principaltypeinfer}. While these approaches are quite accurate, they are limited by the low coverage problem caused by dynamic features and external calls~\cite{hityper}.

\begin{figure}[t]
    \centering
    \includegraphics[width = 0.47 \textwidth]{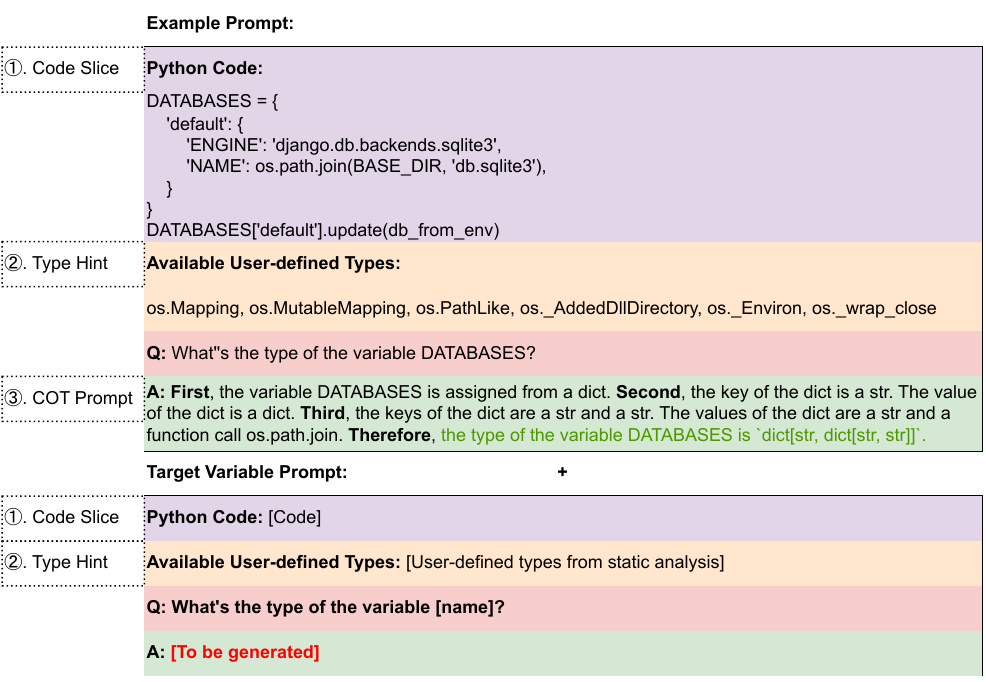}
    \caption{Input prompt with example from the code in Fig.~\ref{fig:code}.}
    \label{fig:prompt}
\end{figure}

\begin{figure*}
    \centering
    \includegraphics[width = 0.85 \textwidth]{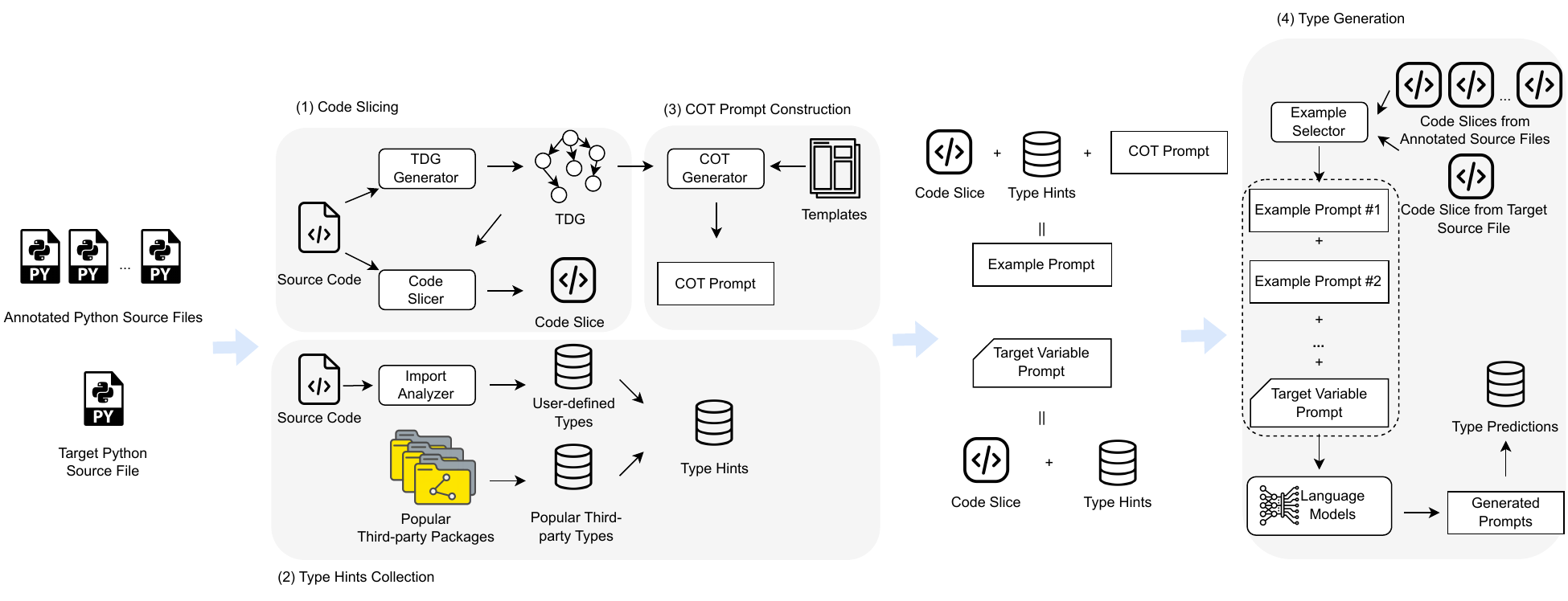}
    \caption{The overview of \tool.}
    \label{fig:overview}
\end{figure*}

\subsection{Supervised Type Inference}

Supervised type inference approaches utilizing deep learning models have made significant progress in predicting types for dynamic languages. Fig.~\ref{fig:background}(2) illustrates the typical process of these approaches: features are extracted from code and encoded into vectors using deep learning models such as recurrent neural networks (RNNs)~\cite{rnn}. A classifier is then used to classify the vectors into pre-defined types. The loss is calculated based on the type prediction of the classifier and the human type annotation, and the parameters of the deep learning models and classifier are updated via back-propagation.

Allamanis \etal~\cite{typilus} adopt an open vocabulary model which encodes code as graphs to predict types. Pradel \etal~\cite{pradel20typewriter} uses multiple RNN models to encode features such as identifiers and code tokens. Mir \etal~\cite{mir2021type4py} improve the top-1 accuracy via a deep similarity clustering algorithm. Wei \etal~\cite{wei20lambdanet} propose to use graph neural networks to predict types. Jesse \etal~\cite{jesse21typebert} propose \typebert by reformulating type prediction as a NER problem. Peng \etal~\cite{peng22hityper} propose HiTyper, which uses deep learning models to recommend types for static inference. While these approaches achieve satisfying performance, they require high-quality datasets for training, which can be difficult to obtain in the wild. Furthermore, supervised type inference approaches only provide predictions without any explanation about how they infer the types, making it challenging for developers to understand and verify the results.

\subsection{Cloze-Style Type Inference}

To enhance the reliability of Python software and prevent potential type errors, the Python Software Foundation has introduced a series of Python Enhancement Proposals (PEPs)\cite{pep484,pep544,pep585,pep589} that enable developers to add static type annotations to their code. As these annotations become part of the code, they can be leveraged by pre-trained code models that are trained on a vast amount of open-source Python programs. Cloze-style type inference approaches, as illustrated in Fig.\ref{fig:background}(3), add masks on the locations of type annotations in the code and invoke pre-trained code models to fill in the masks with predicted types.

All pre-trained code models with Masked Language Modeling (MLM) training objectives such as CodeBERT~\cite{feng2020codebert}, GraphCodeBERT~\cite{guo21graphcodebert} and \codet~\cite{wang21codet5} can be naturally used to predict type annotations. Most recently, \unixcoder~\cite{guo22unixcoder} is a unified cross-modal pre-trained model to support both code-related understanding and generation tasks. \incoder~\cite{daniel22incoder} is a large code generative model that can refill arbitrary regions of code. Leveraging pre-trained code models, cloze-style type inference approaches can be readily implemented. However, they still exhibit limited performance as they solely rely on the general knowledge of pre-trained code models. These approaches can hardly handle complicated types without domain knowledge from static typing rules, and their predictions lack interpretability without explanations.
\section{Generative Type Inference}\label{sec:approach}

\subsection{Overview}

As a generative approach, \tool first generates domain knowledge-aware prompts and then inputs them into language models for type prediction. To achieve this, \tool adopts the widely-used \textit{in-context learning} methodology~\cite{dong2023icl}. This methodology provides a few example questions and answers as demonstrations for the language model and then asks the answer for a new question. Leveraging this methodology, \tool constructs the input prompt by adding some domain knowledge-aware \textit{example prompts} (example questions and answers) before the \textit{target variable prompt} (new question). Fig.~\ref{fig:prompt} illustrates an input prompt with an example from the code in Fig.~\ref{fig:code}. The domain-aware \textit{example prompts} include three parts: code slice, type hint and COT prompt, as shown in Fig.~\ref{fig:prompt}. They are designed to incorporate different static domain knowledge for language models. Specifically, the \textbf{code slice} isolates the statements contributing to the construction of the type for the target variable, with the remaining unrelated statements removed. 
The \textbf{type hint} includes external knowledge that is specific to different code slices, including user-defined types and third-party types. The \textbf{COT prompt} indicates the inference steps of static analysis, aiming at teaching language models how to infer types. The \textit{target variable prompt} contains only the code slice and the type hint of the target variable.

We provide an overview of \tool's workflow in Fig.~\ref{fig:overview}. To start, \tool takes a set of annotated Python source files to select examples and a target Python source file where the target variable is located. For each target source file, \tool generates an input prompt that incorporates domain-aware example prompts and the target variable prompt. A language model is then employed to produce the COT prompt, which includes both the predicted type and corresponding explanations. To generate domain-aware example prompts, \tool conducts three phases: (1) \textit{code slicing}, (2) \textit{type hint collection}, and (3) \textit{COT prompt construction} to generate the code slice, type hint, and COT prompt, respectively. Finally, in the (4) \textit{Type Generation} phase, \tool leverages in-context learning to infer the types of target variables.

\subsection{Code Slicing}
The code slicing phase aims at identifying the code statements related to the target variable based on the type dependency graph (TDG) which indicates the type dependencies among variables~\cite{peng22hityper}.

\subsubsection{TDG Generation} 
\begin{figure}[t]
    \centering
    \includegraphics[width = 0.45 \textwidth]{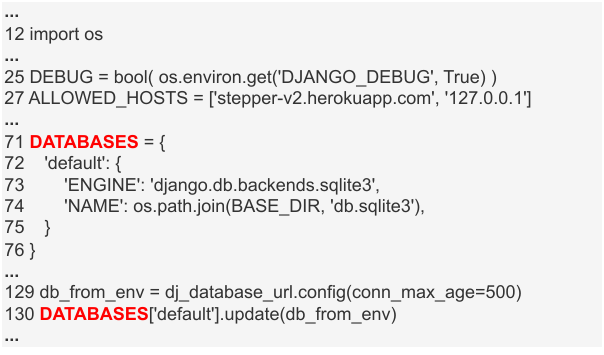}
    \caption{The source code for the example prompt in Fig.~\ref{fig:prompt}, where DATABASES is the target variable.}
    \label{fig:code}
\end{figure}

\begin{figure}[t]
    \centering
    \includegraphics[width = 0.48\textwidth]{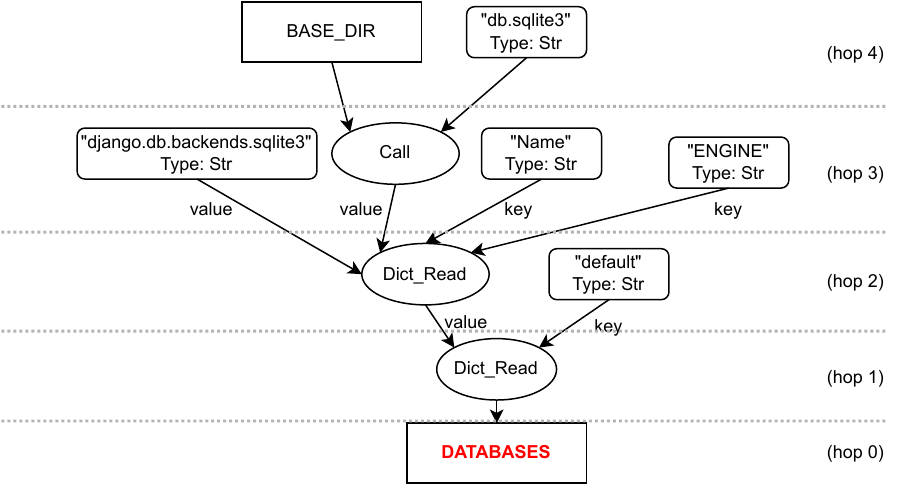}
    \caption{The sliced type dependency graph (TDG) of code in Fig.~\ref{fig:code}.}
    \label{fig:tdg}
\end{figure}

In order to extract the type dependencies of the target variable, \tool generates type dependency graphs (TDGs) using \hityper~\cite{peng22hityper}. A TDG is a directed graph $(N, E)$, where $N$ is the node set and $E$ is the edge set. Each node $n\in N$ in TDG represents a variable (symbol node), an operation (operation node), or a type (type node), while each edge $e \in E$ indicates that the type of the output node depends on the type of the input node. If the target variable is an argument, a return value or a local variable in the function, \tool generates the TDG for the specific function. Otherwise, if the target variable is a global variable, \tool generates the TDG for all statements in the source file except class definitions and function definitions. To better illustrate the code slicing phase, we give a code example in Fig.~\ref{fig:code} and its sliced TDG in Fig.~\ref{fig:tdg}, where the target variable is ``DATABASES''.

To refine the initial TDG, \tool prunes the nodes that do not have any type dependency with the target variable. For instance, in the code presented in Fig.\ref{fig:code}, \tool removes the nodes generated from statements at lines 25, 27, 129, etc. \tool then merges identical symbol nodes that are directly connected, since they represent different occurrences of the same variable. To generate the sliced TDG, \tool locates the sub-graph where the target variable is defined in the refined TDG. For the example in Fig.\ref{fig:code}, \tool identifies the target variable node ``DATABASES'' at line 71 and extracts the reachable sub-graph of the refined TDG as the sliced TDG, as illustrated in Fig.~\ref{fig:tdg}.

\subsubsection{Code Slice Generation}\label{sec:slicing} Using the sliced TDG, \tool generates a code slice from the original input source code, containing only the statements that have type dependencies with the target variable. In this way, \tool reduces the entire function into a smaller code slice that includes only the information relevant to the type inference of the target variable. \tool employs different strategies for generating code slices for local variables, return values, and arguments.

\textbf{Local Variables and Return Values.} To generate code slices for local variables and return values, \tool leverages the clear definitions that indicate how their types are constructed in the code. \tool initiates the process by starting from the definition node of the target variable and traversing \textbf{backward} on the TDG, i.e., in the opposite direction to the TDG edges, to include all the nodes that contribute to the definition of the target variable. The distance of each node from the target variable node is determined by the number of edges (hops) between them. A maximum threshold is also set to prevent infinite loops, and any nodes with distances exceeding the threshold are removed. Once the traversal is completed, \tool combines the corresponding statements of the remaining nodes in the TDG to form the code slice.

\textbf{Arguments.} To facilitate the generation of code slices for function arguments, \tool adopts a different approach, since arguments do not have explicit definitions. 
\tool collects the usages of function arguments, as variable usages often provide hints about their types. For instance, operations such as \textit{open} are usually associated with \textit{File} types. Thus, \tool begins with all nodes of the target argument and traverses \textbf{forward} on the TDG, i.e., in the same direction as the TDG edges, to include nodes that use the target argument. \tool generates code slices for nodes with distances within a specified maximum threshold, similar to local variables and return values.

\subsection{Type Hints Collection}
Unlike built-in types that are available in every source file, user-defined and third-party types are specific to each source file and are defined through class definitions and import statements. However, general knowledge bases in language models do not cover this specific domain knowledge~\cite{peng22hityper}. To address this issue, \tool collects all the user-defined and third-party types that are imported to the current source file as type hints.

To identify user-defined types, \tool performs an import analysis on the current source directory. First, it collects all class definitions in the current source file as user-defined types. Then, it examines the import statements to determine which source files are imported in the current file and adds their class definitions to the list of user-defined types. For third-party types, following previous study~\cite{ye22knowledge}, \tool downloads the top 10,000 popular Python packages ranked by libraries.io~\cite{libio} and employs the same import analysis technique to identify third-party types. All third-party types collected by \tool are stored in a database, which can be queried by \tool to identify the available third-party types based on the import statements in the current source file.

When generating type hints, \tool analyzes all the import statements in the current source file. If a user-defined package is imported, \tool directly conducts import analysis to gather all available user-defined types. If a third-party package is imported, \tool queries the database and obtains all available third-party types. All available types are concatenated to build the type hint, with an example shown in Fig. \ref{fig:prompt} (highlighted in orange color). To prevent excessively long type hints, \tool imposes a maximum threshold (set to 50 in this paper) for the number of collected types. Considering the scarcity of user-defined types, \tool prioritizes the importance of types based on the following order: ``user-defined types in current source file $>$ user-defined types in other source files $>$ third-party types''.

\subsection{Chain-of-Thought Prompt Construction}
\begin{table}[t]
    \centering
    \caption{Chain-of-Thought Prompt Template. [NAME] indicates the name of symbol nodes, [OP] indicates the name of operation nodes and [TYPE] indicates the name of type nodes. [GTTYPE] indicates the annotated type for the target variable. DD-RV and DD-A indicate the dependency description for local variables and return values, and arguments, respectively.}
    \scalebox{0.9}{
    \begin{tabular}{lll}
    \toprule
        \textbf{Part} & \textbf{Type} &  \textbf{Template}\\
    \midrule
        \multirow{10}*{\tabincell{l}{DD-RV}}& Operation$\rightarrow$Symbol & \tabincell{l}{The variable/return value of [NAME] \\ is assigned from [OP] operation.} \\
        &Symbol$\rightarrow$Symbol & \tabincell{l}{The variable/return value of [NAME] \\ is assigned from variable [NAME].}\\
        &Type$\rightarrow$Symbol & \tabincell{l}{The variable/return value of [NAME] \\ is assigned from [TYPE].}\\
        &Operation$\rightarrow$Operation &  \tabincell{l}{The operand(s)/target(s)/key(s)/value(s) \\ of [OP] is/are [OP] operation.} \\
        &Symbol$\rightarrow$Operation & \tabincell{l}{The operand(s)/target(s)/key(s)/value(s) \\ of [OP] is/are variable [NAME].}\\
        &Type$\rightarrow$Operation & \tabincell{l}{The operand(s)/target(s)/key(s)/value(s) \\ of [OP] is/are [TYPE].}\\
        \midrule
        \multirow{4}*{DD-A}&Usage & \tabincell{l}{The argument [NAME] is used \\ in [OP]/[NAME].}\\
        &Naming & \tabincell{l}{Based on the naming convention, \\ it is reasonable to assume that the type of \\ the argument [NAME] is [GTTYPE].} \\
        \midrule
        Con & Conclusion & \tabincell{l}{Therefore, the type of the variable/return \\ value of/argument [NAME] is [GTTYPE].} \\
    \bottomrule
    \end{tabular}}
    \label{tab:cot}
\end{table}

\tool translates the type inference steps of static analysis into chain-of-thought (COT) prompts~\cite{cot} to involve static domain knowledge of how a type is constructed, where a COT
is a series of intermediate reasoning steps~\cite{cot}. To generate COT prompts, \tool utilizes the sliced TDG produced by the \textit{code slicing} phase.

Given the sliced TDG, \tool first organizes the nodes into different hops according to their distance from the target variable node. The hop of the node of the target variable is set to 0, and the hops of other nodes are determined by their distance, as shown in Fig.~\ref{fig:tdg}. The COT prompt constructed by \tool includes \textit{dependency description} and \textit{conclusion}. The \textit{dependency description} explains how the type is inferred, whereas the \textit{conclusion} gives the final type prediction. \tool translates each hop in the TDG into a sentence of \textit{dependency description} and generates the \textit{conclusion} based on the annotated types from developers. Following the recent study~\cite{zhang2022automatic}, we summarize the powerful templates of COT prompts from the zero-shot outputs of language models and present them in Table~\ref{tab:cot}. For the \textit{conclusion}, \tool fills in the variable name and annotated type into the template. Regarding the \textit{dependency description}, \tool employs different prompt templates for local variables, return values, and arguments as \tool utilizes different information for them in the sliced TDGs  in Sec.~\ref{sec:slicing}.

\textbf{Local Variables and Return Values.} Local variables and return values have clear definitions in the code, so it is possible to construct a comprehensive description of how the type of target variable should be inferred. To construct the \textit{dependency description} for them, \tool starts with the edges connected to the target variable node and traverses backward on the TDG to translate each edge into a sentence of \textit{dependency description}. Since there are three major types of nodes in the TDG, we design six templates for six kinds of edges in TDG to generate \textit{dependency descriptions}, as shown in the first part of Table~\ref{tab:cot}. Note that the type node cannot be the output node as its type does not depend on other nodes. Take the TDG in Fig.~\ref{fig:tdg} as an example. There is an edge from operation node \textit{Dict\_Read} to symbol node \textit{DATABASES}. For this edge, \tool adopts the \textit{Operation$\rightarrow$Symbol} template and generates a sentence ``\textit{The variable DATABASES is assigned from a dict}''. There is also an ordinal number at the beginning of each sentence to indicate one inference step. Sentences generated from all edges are concatenated together according to the backward traversal order to form a complete \textit{dependency description}.

\textbf{Arguments.} As arguments usually do not have clear definitions in the code, it is difficult for static analysis to infer their types. Rather than providing solid definition information, \tool provides usage and naming information as hints for type prediction. We present the usage template and naming template in the second part of Table~\ref{tab:cot}. For usage information, \tool collects all nodes in the sliced TDG and constructs the sentence ``\textit{The argument ... is used in ...}''. For the naming information, \tool adds the sentence ``\textit{Based on the naming convention, it is reasonable to assume that the type of the argument is ...}'' to remind language models to consider the argument name. These two sentences form a complete \textit{dependency description} for arguments.

The generated \textit{dependency description} and \textit{conclusion} are finally combined together to form a complete COT prompt. Fig.~\ref{fig:prompt} shows the COT prompt generated by \tool for the code example in Fig.~\ref{fig:code}, highlighted in green color.

\subsection{Type Generation}

For each input source file and target variable, \tool generates its corresponding code slice and selects a set of code slices from the training set as examples based on BM25 similarity~\cite{stephen09the}. BM25 similarity calculates the token similarity between two code slices and has been widely used in recent studies~\cite{li21editsum,gao2023constructing}. Following previous work~\cite{Liu2021WhatMG}, the example prompts are ordered based on the BM25 similarity of code slices: $\{EP_1, ..., EP_n\}$($\forall i \in [1,n)\ BM25(EP_i, TP) \leq BM25(EP_{i+1}, TP)$)  where $EP$ is an example prompt and $TP$ is the target variable prompt. The example prompts are then combined with the target variable prompt to form the complete input prompt, as shown in Fig.~\ref{fig:prompt}.

Given the examples in the input prompt, language models can learn to generate a similar COT prompt for the target variable. To facilitate the automatic evaluation of the generated COT prompts, \tool surrounds type predictions in example COT prompts with quotes, such as \`dict[str, dict[str, str]]\` in Fig.~\ref{fig:prompt}. This allows language models to learn to emphasize type predictions in generated COT prompts by adding quotes. Ultimately, \tool extracts the content within the quotes as the types predicted by the language models.

\section{Experiment Setup}\label{sec:setup}

\subsection{Dataset}

\begin{table}[t]
    \centering
    \caption{The statistics of the ManyTypes4Py dataset. `Arg'' indicates function arguments, ``Ret'' indicates function return values, ``Var'' indicates global and local variables, ``Ele'' indicates elementary types, ``Gen'' indicates generic types, and ``Usr'' indicates user-defined types and third-party types.}
    \scalebox{0.81}{\begin{tabular}{cc|ccc|ccc}
    \toprule
         \textbf{Dataset} & \textbf{Total} & \textbf{Arg} & \textbf{Ret} & \textbf{Var} & \textbf{Ele} & \textbf{Gen} & \textbf{Usr} \\
         \midrule
         \multirow{2}*{\tabincell{c}{Training \\ Set}} & 242,954 & 48,461 & 22,034 & 172,459 & 128,006 & 67,185 & 47,763 \\
         & 100\% & 20.0\% & 9.1\% & 70.9\% & 52.7\% & 27.6\% & 19.7\% \\
         \midrule
         \multirow{2}*{\tabincell{c}{Test
         \\ Set}} &85,205 & 16,700 & 7,754 & 60,751 & 44,605  & 23,310 & 17,290 \\
         & 100\% & 19.6\% & 9.1\% & 71.3\% & 52.5\% & 27.4\% & 20.1\% \\
        \midrule
         \multirow{2}*{\tabincell{c}{Sampled \\ Test
         Set}} & 10,000 & 1,995 & 914 & 7,091 & 5,199  & 2,748 & 2,053 \\
         & 100\% & 20.0\% & 9.1\% & 70.9\% & 52.0\% & 27.5\% & 20.5\% \\
         \bottomrule
         
    \end{tabular}
    }
    \label{tab:dataset}
\end{table}

We follow previous work~\cite{mir2021type4py, peng22hityper} and evaluate our approach on the ManyTypes4Py dataset~\cite{type4pydataset} by splitting the dataset into a training set and a test set with an 80:20 ratio. We use the training set to train the baseline models and select examples for \tool and evaluate the performance of \tool and baselines in the test set. In order to accommodate the computational resource limitations, we further sample 10,000 instances from the test set during the evaluation of large language models. Table~\ref{tab:dataset} presents the statistics of the experimental datasets.

\subsection{Baselines}

We choose the following four type inference approaches as our baselines:
\begin{itemize}
    \item \textbf{\typebert}~\cite{jesse21typebert} is a supervised type inference apporoach. It reformulates the type inference problem into a Named Entity Recognition (NER) problem and regards types as labels.
    \item \textbf{\typewriter}~\cite{pradel20typewriter} is a supervised type inference approach. It extracts different code features, such as identifiers and code tokens, and utilizes four RNNs to encode the extracted features and make type predictions.
    \item \textbf{\typepy}~\cite{mir2021type4py} is a supervised type inference approach. It builds different type clusters and classifies a new Python program into one of the type clusters to determine the type.
    \item \textbf{\hityper}~\cite{peng22hityper} is a hybrid type inference approach. It builds a type dependency graph (TDG) for each target variable and utilizes both static analysis and neural prediction to fill the blanks in the TDG and finally outputs the validated types.
\end{itemize}

We choose three popular pre-trained code models, including
\codet~\cite{wang21codet5}, \unixcoder~\cite{guo22unixcoder}, and \incoder~\cite{daniel22incoder} to represent the performance of cloze-style type inference approaches. Besides, we choose \gptneo~\cite{gptneo}, \gptj~\cite{gptj}, \codegen~\cite{nijkamp2023codegen}, \gpt~\cite{brown2020GPT3} and \chatgpt~\cite{chatgpt} to evaluate the performance of \tool on language models with different parameter sizes. We present the statistics of all the language models in Table~\ref{tab:model}.

\begin{table}[t]
    \centering
    \caption{The statistics of language models used in the evaluation.}
    \scalebox{0.9}{
    \begin{tabular}{lccc}
    \toprule
       \textbf{Model}  &  \textbf{\#Parameters} & \textbf{Training Dataset} & \textbf{Type} \\
    \midrule
        CodeT5 & 220M/770M & \tabincell{c}{CodeSearchNet \\ \& BigQuery} & \tabincell{c}{Generative \\ \& Infilling}\\
        \g UniXcoder & \g 126M & \g CodeSearchNet & \g Infilling\\
        GPT-Neo &  1.3B/2.7B & The Pile & Generative \\
        \g InCoder & \g 1.3B/6.7B & \g \tabincell{c}{GitHub \& \\ Stack Overflow} & \g \tabincell{c}{Generative \\ \& Infilling}\\
        CodeGen & 6B & \tabincell{c}{The Pile, BigQuery \\ \& GitHub} & Generative \\
        \g GPT-J   &  \g 6.7B & \g The Pile & \g Generative \\
        GPT-3.5 & 175B & - & Generative \\
        \g ChatGPT & \g 175B & \g - & \g Generative \\
    \bottomrule
    \end{tabular}}
    \label{tab:model}
\end{table}

\subsection{Metrics}
We use two commonly used metrics in previous work~\cite{typilus,mir2021type4py,peng22hityper} to evaluate the performance of \tool and other baselines:
\begin{itemize}
    \item \textbf{Exact Match} is defined by the ratio of type predictions made by an approach that \textbf{exactly} match type annotations from developers. 
    \item \textbf{Match to Parametric} is defined by the ratio of type predictions made by an approach that \textbf{share the common outmost type} with type annotations from developers.
\end{itemize}
For example, \textit{List[int]} and \textit{List[str]} are considered Match to Parametric but not Exact Match since they are different types while sharing the same outmost type \textit{List}.

\subsection{Implementation}
For the four type inference baselines \typebert~\cite{jesse21typebert}, \typewriter~\cite{pradel20typewriter}, \typepy~\cite{mir2021type4py} and \hityper~\cite{peng22hityper}, we directly use the replication packages released by the authors and other researchers. For all the language models except \gpt and \chatgpt, we download them from HuggingFace Hub~\cite{huggingface} and deploy them locally. Following the previous work~\cite{mir2021type4py,peng22hityper}, we adopt the generated sentences with top-5 probabilities as predictions. For \gpt and \chatgpt, we use the public APIs provided by OpenAI under engine ``\textit{text-davinci-003}'' and ``\textit{gpt-3.5-turbo-0301}'', respectively. We 
acquire 50 samples with a temperature of 1.0 for each target variable and rank the top-5 predictions according to the occurrence frequency, following the work~\cite{xia22practical, xia23conversational}. We choose the maximum distance threshold of TDG nodes at 3 for the code slicing phase of \tool, as previous studies~\cite{typilus,mir2021type4py} only consider types with nested levels smaller than 3. All experiments are conducted on a Linux machine (Ubuntu 18.04) with one 112-core Intel Xeon Gold 6348 CPU@ 2.60GHz, two NVIDIA A100-80GB GPUs, and 1TB RAM.
\section{Evaluation}\label{sec:eval}

\begin{table*}[t]
    \centering
    \caption{The performance of \tool along with the baselines under four types of variables in terms of Top-1,3,5 Exact Match (\%) and Match to Parametric (\%). ``Arg'', ``Ret'', ``Var'', and ``All'' indicate function arguments, function return values, global and local variables, and all of above, respectively.
    Under each metric the best performance is marked as \colorbox{mygray}{gray}.}
    \scalebox{0.9}{
    \begin{tabular}{cclcccccccccccccccc}
    \toprule
        \multirow{2}*{\textbf{Metric}} & \multirow{2}*{\textbf{Category}}  &\multirow{2}*{\textbf{Approach}} & & \multicolumn{4}{c}{\textbf{Top-1}}  & & \multicolumn{4}{c}{\textbf{Top-3}} &  & \multicolumn{4}{c}{\textbf{Top-5}}  \\
        \cmidrule{5-8}
        \cmidrule{10-13}
        \cmidrule{15-18}
         & & & & Arg & Ret & Var & All & & Arg & Ret & Var & All & & Arg & Ret & Var & All \\
        \midrule
         \multirow{10}*{\tabincell{c}{\textbf{Exact} \\ \textbf{Match} \\ (\%) }} & \multirow{3}*{\tabincell{c}{Supervised}} &  \typebert & & 28.0 & 38.5 & 51.1 & 45.4 & & 34.8 & 52.6 & 55.8 & 51.4 & & 36.5 & 57.1 & 58.6 & 54.1\\
        \specialrule{0em}{1pt}{1pt}
        & &\typewriter& & 53.3 & 52.8 & - & - & & 61.1 & 60.7 & - & - & & 65.8 & 65.3 & - & -\\
        \specialrule{0em}{1pt}{1pt}
        & &\typepy& & 66.5 & 56.1 & 82.0 & 76.6 & & 72.0 & 59.2 & 83.8 & 79.3 & & 73.8 & 60.7 & 84.3 & 80.1\\
        \specialrule{0em}{1pt}{1pt}
        \cmidrule{2-18}
        & \multirow{5}*{\tabincell{c}{Cloze \\ Style}} & \incoder-1.3B & & 20.9 & 20.5 & 15.1 & 16.7 & & 21.3 & 20.8 & 15.5 & 17.1 & & 21.3 & 21.0 & 15.6 & 17.2\\
        \specialrule{0em}{1pt}{1pt}
        &&\incoder-6.7B && 24.1 & 42.0 & 18.7 & 21.9 && 24.6 & 42.7 & 19.1 & 22.3 && 24.7 & 43.1 & 19.2 & 22.4 \\
        \specialrule{0em}{1pt}{1pt}
        & &\unixcoder & & 55.0 & 49.2 & 35.9 & 40.9 & & 66.9 & 64.6 & 42.1 & 49.0 & & 70.6 & 69.8 & 45.2 & 52.4\\
        \specialrule{0em}{1pt}{1pt}
        & &\codet-base & & 51.1 & 57.6 & 21.7 & 30.7 & & 59.3 & 64.4 & 28.0 & 37.4 & & 62.0 & 66.9 & 30.7 & 40.1\\
        \specialrule{0em}{1pt}{1pt}
        & &\codet-large & & 56.2 & 60.2 & 44.7 & 48.4 & & 61.6 & 64.5 & 50.4 & 53.9 & & 63.9 & 66.3 & 53.4 & 56.6\\
        \cmidrule{2-18}
        & Generative &\tool& & \g 73.1 & \g 68.7 & \g 82.2 & \g 79.2 & & \g 81.0 & \g 77.1 & \g 87.9 & \g 85.6 & & \g 82.7 & \g 79.1 & \g 89.1 & \g 87.0\\
        \cmidrule{1-18} 
        \multirow{10}*{\tabincell{c}{\textbf{Match to} \\ \textbf{Parametric} \\ (\%)}} & \multirow{3}*{\tabincell{c}{Supervised}} &  \typebert & & 29.8 & 41.4 & 54.0 & 48.1 & & 36.0 & 55.9 & 58.0 & 53.5 & & 37.7 & 60.8 & 61.2 & 56.5\\
        \specialrule{0em}{1pt}{1pt}
        & &\typewriter& & 54.4 & 54.1 & - & - & & 63.4 & 63.5 & - & - & & 68.8 & 69.3 & - & -\\
        \specialrule{0em}{1pt}{1pt}
        & &\typepy& & 68.0 & 59.0 & 86.2 & 80.2 & & 74.1 & 64.1 & 88.3 & 83.3 & & 75.9 & 66.3 & 88.8 & 84.3\\
        \specialrule{0em}{1pt}{1pt}
        \cmidrule{2-18}
        & \multirow{5}*{\tabincell{c}{Cloze \\ Style}} & \incoder-1.3B & & 22.9 & 22.8 & 18.7 & 19.9 & & 23.3 & 23.1 & 19.1 & 20.3 & & 23.4 & 23.3 & 19.2 & 20.4\\
        \specialrule{0em}{1pt}{1pt}
        & &\incoder-6.7B & & 28.8 & 51.6 & 25.0 & 28.1 & & 29.3 & 52.1 & 25.3 & 28.5 & & 29.4 & 52.5 & 25.3 & 28.6\\
        \specialrule{0em}{1pt}{1pt}
        & &\unixcoder & & 61.9 & 61.8 & 44.3 & 49.3 & & 72.3 & 76.0 & 51.2 & 57.6 & & 75.0 & 80.1 & 53.8 & 60.4\\
        \specialrule{0em}{1pt}{1pt}
        & &\codet-base & & 54.8 & 66.7 & 27.7 & 36.6 & & 62.9 & 74.2 & 34.4 & 43.6 & & 65.6 & 76.4 & 37.1 & 46.3\\
        \specialrule{0em}{1pt}{1pt}
        & &\codet-large & & 61.4 & 69.4 & 55.7 & 58.0 & & 66.8 & 74.3 & 61.2 & 63.5 & & 68.9 & 76.2 & 63.7 & 65.9\\
        \cmidrule{2-18}
        & Generative &\tool& & \g 78.7 & \g 75.6 & \g 91.2 & \g 87.3 & & \g 84.9 & \g 83.0 & \g 93.7 & \g 91.0 & & \g 86.1 & \g 84.5 & \g 94.1 & \g 91.7\\
    \bottomrule
    \end{tabular}}
    \label{tab:mainres}
\end{table*}

\subsection{Research Questions}
In the evaluation, we focus on the following four research questions:
\begin{itemize}
    \item \textbf{RQ1}: How effective is \tool in type inference compared with existing approaches?
    \item \textbf{RQ2}: How capable is \tool in  language models with different parameter sizes?
    \item \textbf{RQ3}: What are the impacts of different parts in the prompt design of \tool?
    \item \textbf{RQ4}: What are the impacts of different examples in \tool?
\end{itemize}

To study RQ1, we conduct experiments on both \tool and baseline approaches with the entire test set (85,205 instances), aiming to comprehensively verify the effectiveness of \tool against state-of-the-art type inference techniques. However, due to limited computational resources, we only use the sampled test set (10,000 instances) for RQ2-4. For RQ2, we evaluate six language models under \tool to examine the tool's effectiveness across language models with different parameter sizes. We also include two additional settings: \textbf{Zero-Shot} and \textbf{Standard ICL}. In the \textbf{Zero-Shot} setting, we do not provide any example and only use the source code of the target variable as the input prompt for language models in the type prediction. The Zero-Shot setting tests the basic performance of language models on type prediction. In the \textbf{Standard ICL} setting, we provide three fixed example prompts before the target variable prompt and use only source code in the input prompts, which is the same with recent study~\cite{li23towards}. The Standard ICL setting indicates the basic performance of language models with the \textit{in-context learning} methodology. To fairly compare \tool with the Standard ICL setting, we also set the number of examples in \tool to 3 in RQ2. For RQ3, we remove different parts of the prompt design in \tool to study the impacts of each part. For RQ4, we vary the number of examples and the example selection method to investigate the impacts of different examples. We choose \chatgpt as the base model of \tool and use five examples in \tool in all the RQs except RQ2.

\subsection{RQ1: Effectiveness of \tool}

\begin{table}[t]
    \centering
    \caption{The performance of \hityper with different base models under four types of variables. Variable categories are the same with Table~\ref{tab:mainres}.}
    \scalebox{0.9}{
    \begin{tabular}{clcccc}
    \toprule
        \textbf{Metric} & \textbf{Base Model}  & \textbf{Arg} & \textbf{Ret} & \textbf{Var} & \textbf{All} \\
        \midrule
         \multirow{3}*{\tabincell{c}{\textbf{Exact} \\ \textbf{Match} \\ (\%)}} & - & 8.0 & 43.5 & 65.7 & 52.4\\
         \specialrule{0em}{1pt}{1pt}
        & \typepy &  73.5 & 73.4 & 90.6 & 85.7\\
        \specialrule{0em}{1pt}{1pt}
         & \tool &  84.9 & 77.9 & 90.5 & 88.3\\
        \midrule
        \multirow{3}*{\tabincell{c}{\textbf{Match to} \\ \textbf{Parametric} \\ (\%)}} & - & 8.4 & 52.7 & 70.2 & 56.5\\
        \specialrule{0em}{1pt}{1pt}
        & \typepy & 76.1 & 83.4 & 95.3 & 90.1\\
        \specialrule{0em}{1pt}{1pt}
        & \tool &  87.4 & 87.3 & 95.3 & 93.1\\
    \bottomrule
    \end{tabular}}
    \label{tab:hybrid}
\end{table}

\subsubsection{Comparison with Supervised Approaches} We compare \tool with three supervised type inference approaches, namely \typebert, \typewriter, and \typepy. The results are presented in Table~\ref{tab:mainres}, where we report the top-1, top-3, and top-5 Exact Match and Match to Parametric for four categories of variables. It is worth noting that \typewriter is designed solely for argument and return value type predictions. Analyzing the top-1 prediction results in Table~\ref{tab:mainres}, we observe that \tool outperforms the best supervised approach, \typepy, by 10.0\% for argument type prediction and 22.5\% for return value type prediction in terms of Exact Match. This improvement of \tool over \typepy is even more significant for top-5 predictions, where \tool outperforms \typepy by 12.1\% for argument type prediction and 30.3\% for return type prediction in terms of Exact Match. Moreover, when considering Match to Parametric, \tool achieves a consistent improvement of 8.7\% $\sim$ 9.3\% on overall variables than \typepy. These results demonstrate that, even with few annotated examples, the generative type inference approach \tool is more effective than supervised approaches such as \typepy. We also observe that \tool does not perform much better than \typepy on predicting local variables. This can be attributed to the lower difficulty of type inference for local variables compared to arguments and return values, so static domain knowledge incorporated by \tool provides limited improvements. This can also be verified by Table~\ref{tab:mainres}, where all the supervised approaches obtained much higher performance on local variables than arguments and return values.

\subsubsection{Comparison with Cloze-Style Approaches} We compare \tool with cloze-style approaches, namely \incoder, \unixcoder and \codet, and present the results in Table~\ref{tab:mainres}. Our observations indicate that, in general, cloze-style approaches perform worse than supervised approaches due to their lack of domain knowledge from data and static analysis. By introducing five annotated examples and incorporating static knowledge, \tool outperforms the best cloze-style approach \codet-large by 63.6\% on overall top-1 Exact Match and 53.7\% on overall top-5 Exact Match. This suggests that incorporating domain knowledge from static analysis with a few examples can largely improve the performance of type inference. 


\subsubsection{Comparison in Hybrid Approach \hityper} As \hityper is a hybrid approach, we study its performance with the best supervised approach \typepy and \tool, and present the experiment results in Table~\ref{tab:hybrid}. For \typepy and \tool, we use their top-5 predictions as type recommendations since \hityper can reject wrong types. The results show that \hityper performs poorly when there is no base model, particularly in argument type inference, where it achieves an Exact Match of only 8\%. This verifies the low coverage problem of static analysis. When associating with base models, \hityper with \tool still outperforms \hityper with \typepy by 15.5\% for argument type inference and 6.1\% for return value inference. This indicates that the performance gap of \tool over \typepy cannot be bridged by simply combining them with static analysis.

\answer{1}{\tool outperforms the best baseline \typepy by 8.6\% on all variables, with particularly notable improvements of 12.1\% and 30.3\% for argument and return value types, respectively, in terms of top-5 Exact Match.}

\begin{table}[t]
    \centering
    \caption{The performance of different language models under three settings for all variables in terms of Top-1,3,5 Exact Match (\%). $\bigtriangleup$ indicates the improvement of Standard ICL and \tool over the Zero-Shot setting.}
    \scalebox{0.9}{
    \begin{tabular}{cclll}
    \toprule
        \textbf{Base Model}  &\textbf{Approach} & \textbf{Top-1 ($\bigtriangleup$)} & \textbf{Top-3 ($\bigtriangleup$)} & \textbf{Top-5 ($\bigtriangleup$)}  \\
        \midrule
         \multirow{3}*{\tabincell{c}{GPT-Neo \\ (1.3B)}} &  Zero-Shot & 31.5 & 40.6 & 42.8\\
        \specialrule{0em}{1pt}{1pt}
        &Standard ICL & 44.0 (40\%) & 50.0 (23\%) & 50.8 (19\%)\\
        \specialrule{0em}{1pt}{1pt}
        &\tool & 57.0 (81\%) & 61.5 (51\%) & 62.8 (47\%)\\
        \midrule
        \multirow{3}*{\tabincell{c}{GPT-Neo \\ (2.7B)}} &  Zero-Shot& 43.2 & 50.0 & 51.9\\
        \specialrule{0em}{1pt}{1pt}
        &Standard ICL & 46.6 (8\%) & 52.3 (5\%) & 52.8 (2\%)\\
        \specialrule{0em}{1pt}{1pt}
        &\tool & 55.5 (28\%) & 61.9 (24\%) & 63.0 (21\%)\\
        \midrule
        \multirow{3}*{\tabincell{c}{\gptj \\ (6.7B)}} &  Zero-Shot & 42.4  & 43.7 & 43.9\\
        \specialrule{0em}{1pt}{1pt}
        &Standard ICL  & 50.8 (20\%)  & 54.9 (26\%)  & 55.3 (26\%)\\
        \specialrule{0em}{1pt}{1pt}
        &\tool& 62.7 (48\%) & 67.3 (54\%) & 68.4 (56\%)\\
        \midrule
        \multirow{3}*{\tabincell{c}{\codegen \\ (6B)}} &  Zero-Shot & 34.7 & 44.0 & 45.5\\
        \specialrule{0em}{1pt}{1pt}
        &Standard ICL & 54.1 (56\%) & 60.5 (38\%) & 61.9 (36\%)\\
        \specialrule{0em}{1pt}{1pt}
        &\tool & 63.7 (84\%) & 69.1 (57\%) & 70.8 (56\%)\\
        \midrule
        \multirow{3}*{\tabincell{c}{\gpt \\ (175B)}} &  Zero-Shot & 62.0 & 65.4 & 66.3\\
        \specialrule{0em}{1pt}{1pt}
        &Standard ICL & 69.7 (12\%) & 74.2 (13\%) & 75.8 (14\%)\\
        \specialrule{0em}{1pt}{1pt}
        &\tool &  78.9 (27\%) & 85.0 (30\%)  & 86.2 (30\%)\\
        \midrule
        \multirow{3}*{\tabincell{c}{\chatgpt \\ (175B)}} &  Zero-Shot & 61.3 & 66.1 & 67.5\\
        \specialrule{0em}{1pt}{1pt}
        &Standard ICL & 68.0 (11\%)  & 71.8 (9\%)  & 73.1 (8\%)\\
        \specialrule{0em}{1pt}{1pt}
        &\tool & 78.8 (29\%) &  85.3 (29\%) &  86.7 (28\%)\\
    \bottomrule
    \end{tabular}}
    \label{tab:rq2}
\end{table}

\subsection{RQ2: Capability of \tool in Different Language Models}
We compare the performance of \tool on six language models with parameter sizes ranging from 1.3B to 175B with the Zero-Shot setting and the Standard ICL setting and present the overall top-1,3,5 Exact Match in Table~\ref{tab:model}. In the Zero-Shot setting, our results indicate that language models with larger model sizes generally perform better, with \chatgpt achieving a 2$\times$ top-1 Exact Match than \gptneo-1.3B. When providing language models with three fixed examples in the Standard ICL setting, we observe an 8\% $\sim$ 56\% improvement in top-1 Exact Match, demonstrating the effectiveness of the \textit{in-context learning} methodology. For \tool, we find consistent improvements of 27\% $\sim$ 84\% on different language models, with the improvements being more significant for smaller language models like \gptneo-1.3B than larger language models like \chatgpt. With less general knowledge stored in the models, smaller language models benefit more from the domain knowledge associated by \tool. Furthermore, the improvements achieved by \tool over the Zero-Shot setting are $2\times \sim 3\times$ of that achieved by the Standard ICL setting, in terms of top-1 Exact Match. For the top-5 type prediction, \tool even achieves a $10\times$ of improvement obtained by the Standard ICL setting on \gptneo-2.7B. These findings demonstrate the usefulness of incorporating static domain knowledge in the prompt design of \tool, which cannot be outweighed by simply providing some examples.

\answer{2}{\tool is capable of consistently improving the zero-shot performance of type inference for language models with different parameter sizes and achieves $2\times \sim 3\times$ of improvements made by the Standard ICL setting.}

\subsection{RQ3: Impacts of Different Parts of Prompt Design}

To investigate the impact of different parts of the prompt design of \tool, we conduct an ablation study and present the results in Table~\ref{tab:rq3}. The results show that removing code slicing techniques and inputting the whole function of target variables in the prompts leads to a significant performance drop of 24\% on overall type inference. This decrease is mainly caused by local variables and arguments, as there is typically only a small set of statements in the function that have type dependencies with them, while inputting the entire function can introduce useless information and bias the language model. When type hints are removed, the performance of \tool on user-defined types decreases the most (11\%), indicating the importance of providing available user-defined types as additional knowledge for language models. Additionally, when COT prompts are removed, the performance of \tool on generic types drops the most (10\%), as generic types usually involve complicated type dependencies that should be well handled by static analysis. Providing the inference steps of static analysis in COT prompts can greatly help improve the performance of language models on generic types.

\answer{3}{In the prompt design, code slicing improves the overall performance of type inference by 24\%, type hints improve the performance of user-defined type inference by 11\%, and COT prompts improve the performance of generic type inference by 10\%.}

\begin{table}[t]
    \centering
\caption{The performance of \tool when removing different parts of the prompt design in \tool in terms of Top-5 Exact Match (\%). ``Ele'', ``Gen'', and ``Usr'' indicate elementary types, generic types and user-defined types as well as third-party types, respectively. Other variable categories are the same with Table~\ref{tab:mainres}.}
\scalebox{0.9}{
    \begin{tabular}{lccc|ccc|c}
    \toprule
        \textbf{Ablation} &  \textbf{Arg }& \textbf{Ret} & \textbf{Var}  &  \textbf{Ele} & \textbf{Gen} & \textbf{Usr} &  \textbf{All}  \\
        \midrule
         w/o Code Slice  &  74.8 & 77.0 & 68.8 &  75.1 & 75.5 & 73.9 &  70.8\\
        \specialrule{0em}{1pt}{1pt}
        \tabincell{c}{w/o Type Hint}    &  76.1 & 75.9 & 89.3 &  94.1 & 77.2 & 75.9 &  85.5\\
        \specialrule{0em}{1pt}{1pt}
        w/o COT Prompt    &  82.3 & 78.6 & 86.4 &  92.9 & 70.8 & 84.3 &  84.9\\
        \midrule
        \tool   &  83.5 & 79.4 & 89.7 &  94.3 & 77.8 & 84.6 &  87.5\\
    \bottomrule
    \end{tabular}}
    \label{tab:rq3}
\end{table}

\begin{figure}
    \centering
    \includegraphics[width = 0.37\textwidth]{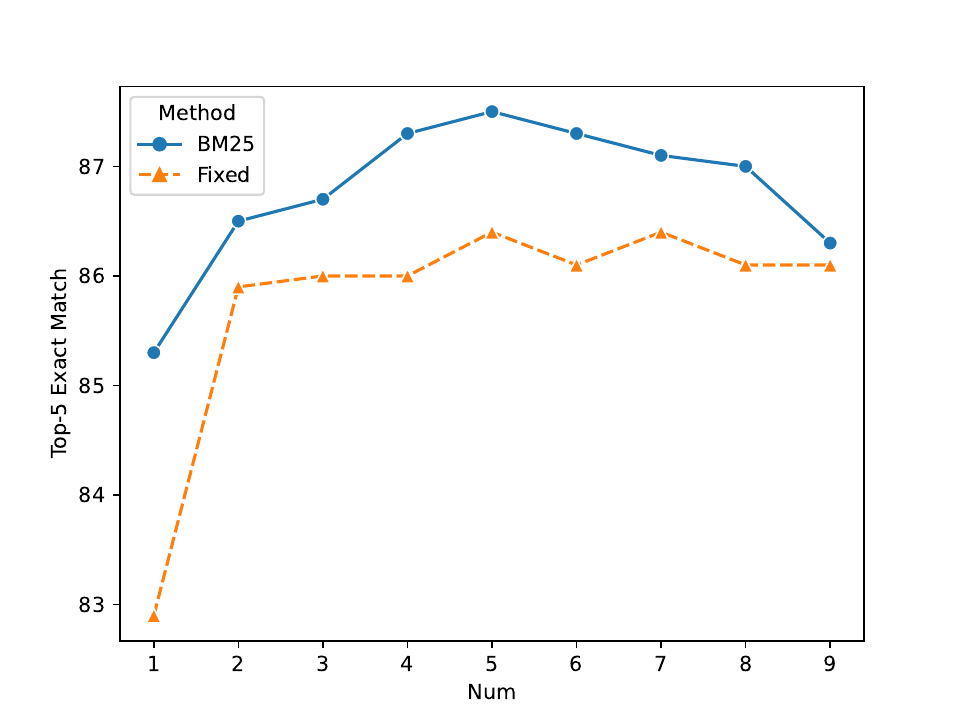}
    \caption{The top-5 Exact Match of \tool with different numbers of examples and different example selection methods}
    \label{fig:num}
\end{figure}

\subsection{RQ4: Impacts of Different Examples}

To evaluate the effects of the number of examples and example selection methods in the prompt design of \tool, we vary the number of examples from one to nine and compare two example selection methods: fixed examples and BM25 similarity-based examples. We present the top-5 Exact Match results of \tool in Fig.~\ref{fig:num}.

Based on the results presented in Fig.~\ref{fig:num}, we observe that the performance of \tool is largely affected by the number of examples provided in the input prompts. Specifically, the performance drops notably when there is only one example, highlighting the importance of providing sufficient examples for effective \textit{in-context learning}. Additionally, the performance of \tool using BM25 similarity-based examples increases up to five examples, after which it starts decreasing. This suggests that both inadequate and excessively long input contexts can harm the performance of \tool. When changing the example selection method, we find that using BM25 similarity-based examples performs better than using fixed examples, particularly when only one example is provided. However, when we provide five examples, the performance drop with fixed examples is relatively small (less than 1.3\%). One possible explanation is that large language models learn how to perform type inference from the example prompts rather than solely relying on the direct correlations between type predictions in different example prompts~\cite{min2022rethinking}.

\answer{4}{\tool achieves the best performance with five examples. \tool shows only a small performance drop ($<$1.3\%) even when provided fixed examples, releasing the burden of developers to design examples. }

\section{Discussion}\label{sec:discussion}
\begin{figure}[t]
    \centering
    \includegraphics[width = 0.4 \textwidth]{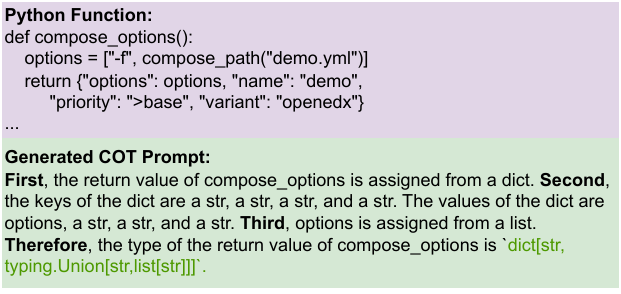}
    \caption{A function whose return value type can only be inferred by \tool. The type annotation for the return value is Dict[str, Union[str, List[str]]].}
    \label{fig:case}
\end{figure}

\subsection{Interpretability of \tool}
To better illustrate the interpretability of COT prompts generated by \tool, we give an example of a function whose return value can only be inferred by \tool in Fig.~\ref{fig:case}. Due to the page limitation, we only present the code slice and the generated COT prompt. To infer the return type of function \textit{compose\_options}, \tool follows similar inference steps as static analysis. First, it infers that the return value is assigned from a dictionary in the generated COT prompt. Then it identifies the types of keys and values by specifying that there are four keys in the dictionary with types of \textit{str}, and there are three values with types of \textit{str} as well as one variable named \textit{options}. The second step in the generated COT precisely matches the code given in the input prompt, indicating that large language models like \chatgpt have the capacity to simulate the inference steps of static analysis. In the third step of the generated COT prompt, \tool recognizes the unknown variable \textit{options} and locates its assignment from a list. Since we set the maximum number of hops to 3, \tool generates the conclusion directly after the third step. From this example, we can find that by providing an explanation in the COT prompt, human developers can easily understand the predictions and determine whether the predictions are correct based on the explanations.

\subsection{Limitations of \tool}

Despite the effectiveness of \tool, we also identify the following limitations:

1) \textit{Limited context.} Although \tool adopts static code slicing techniques based on TDGs, we have observed a limited number of instances in the test set ($\sim$ 1000) with code slices exceeding the maximum context length of language models. This primarily occurs in extremely long functions with complex type dependencies. We recognize that extracting code slices without sacrificing key dependency information is still a challenge.

2) \textit{Limited knowledge for function arguments.} As function arguments lack precise definitions, \tool provides naming and usage information to enable language models to predict their types. However, this information is incomplete and can potentially introduce biases in the model's predictions~\cite{gao2023sides}. Incorporating data flow information via inter-procedural analysis may be a possible solution to enhance argument information.
\section{Conclusion}\label{sec:conclusion}

This paper presents \tool, a few-shot generative type inference method for Python programs. Our approach incorporates static domain knowledge into language models via a novel prompt design in the \textit{in-context} learning paradigm. Experimental results show that \tool outperforms both state-of-the-art supervised type inference methods and cloze-style type inference methods.

\section{Acknowledgement}
The authors would like to thank the efforts made by anonymous reviewers. The work described in this paper was supported by the Research Grants Council of the Hong Kong Special Administrative Region, China (No. CUHK 14206921 of the General Research Fund). The work was also supported by National Natural Science Foundation of China under project (No. 62002084), Natural Science Foundation of Guangdong Province (Project No. 2023A1515011959), Shenzhen Basic Research (General Project No. JCYJ20220531095214031), and Key Program of Fundamental Research from Shenzhen Science and Technology Innovation Commission (Project No. JCYJ20200109113403826). Any opinions, findings, and conclusions or recommendations expressed in this publication are those of the authors, and do not necessarily reflect the views of the above sponsoring entities.

\section{Data Availability}
You can find the code and data related to this paper at https://github.com/JohnnyPeng18/TypeGen.

\bibliographystyle{plain}
\bibliography{ref}

\end{document}